\documentclass[twocolumn]{article} 

\usepackage{authblk}

\usepackage[utf8]{inputenc}
\usepackage{graphicx}
\usepackage{hyperref}
\usepackage{url}
\newcommand{\COO}{CO$_2$ }
\newcommand{\smm}{$\text{mm}^2$}

\usepackage{subcaption}
\usepackage{amsmath}
\usepackage{amsfonts}
\usepackage{xcolor}
\usepackage[]{units}
\usepackage{siunitx}
\usepackage{pgfplots}
\pgfplotsset{compat=1.17}
\usepackage{tikz}
\usetikzlibrary{positioning}
\usetikzlibrary{calc}
\usepackage{arydshln}

\usepackage{multirow}

\usepackage{cleveref}   
\crefname{equation}{Equation}{equations}
\crefname{figure}{Figure}{figures}
\crefname{subsection}{Subsection}{subsections}

\title{The Low Emission Oil\&Gas Open (LEOGO)\\ Reference Platform -- an Off-Grid Energy System\\ for Renewable Integration Studies}
\date{\today}
\author{Harald G Svendsen}
\author{Til Kristian Vrana}
\author{Andrzej Holdyk}
\author{Heiner Sch\"umann}
\affil{SINTEF, Trondheim, Norway}

\begin{document}
\maketitle

\begin{abstract}
This article introduces and describes the integrated energy system of a the Low Emission Oil\&Gas Open (LEOGO) reference platform. It is a hypothetical case meant to represent a typical oil\&gas installation in the North Sea. The aim of this detailed specification is to serve as an open reference case where all information about it can be publicly shared, facilitating benchmarking and collaboration. The relevance of this reference case of an off-grid energy system is not limited to the oil\&gas industry, since it can also been seen as a special kind of electrical micro grid. The remote offshore location makes it especially relevant for studying offshore wind power and ocean energy sources like wave power.

The specification has an emphasis on the energy system and electrical configuration, but includes also a basic description of oil field and processing system. 
The intention is that it will serve as a basis for energy system studies and relating power system stability analyses regarding the integration of renewable energy sources.  
This allows for comparisons of a base case with different design modifications, new operational planning methods, power management strategies and control concepts.
Examples of possible modifications are the replacement of gas turbines by wind turbines, addition of energy storage systems, a more variable operation of loads, etc.

The last part of the article demonstrates the behaviour of the reference platform implemented in two software tools. One for operational planning and one for dynamic power system analyses.

\end{abstract}


\section{Introduction}

To eliminate or reduce \COO emissions from the local energy system at offshore oil\&gas platforms, fossil-fuel based gas turbines must either be equipped with carbon capture and storage technologies~\cite{cepong}, or be replaced by clean alternatives such as
power from offshore wind turbines, power via cable from shore, a shift to hydrogen-based gas turbine fuels or fuel cells, power from other renewable sources, or other solutions that eliminate emissions.
With new energy supply alternatives, new operating strategies are required in order to best utilise the available resources within the given constraints. And the new technology will have different electrical characteristics that must be addressed. 
To demonstrate the adequacy and benefit of new low-emission solutions, new models and methods are needed to analyse system operation and to show that the energy supply system can provide the required security of supply and power system stability.

\subsection{LEOGO reference platform}

To more easily demonstrate and test new operating strategies, different system configurations, control concepts and new ideas and methods for integrating renewables intro an off-grid system, it is useful to have an open and well-defined test case as a common basis for analyses. Therefore, we have created what we call the \emph{Low Emission Oil\&Gas Open} (LEOGO) reference platform. It is intended as a well-defined study case where all underlying data can be publicly shared.  
It constitutes both a \emph{specification} and a \emph{dataset} that is publicly available \cite{leogodata}.

This work has been done within the \emph{LowEmission} research centre, which is a collaboration between research and industry partners aiming to develop technology to reduce emissions from petroleum activities on the Norwegian continental shelf (NCS). The LEOGO specification has been created with crucial input from industry partners in this centre.

Real oil\&gas platforms of course differ from each other in important ways and analyses of this reference platform can never substitute case-specific analyses. 
But often, especially in early phases of development, it is interesting to explore ideas in a general context without a specific case in mind. And from a research perspective, the main interest is often the demonstration of generic concepts where the peculiarities of the study case are less important. Hence we believe that a completely open and freely available case specification can prove useful both for academic research and for industry development.

When required for a particular purpose, the reference case presented here may be adjusted. In such cases, we would encourage authors to indicate very clearly what modifications have been made when reporting their results.

The LEOGO specification in its present form is primarily concerned with the energy system and its links with the topside processing system. Details regarding reservoir and fluid transport from the reservoir to the platform is not included: The boundary is taken to be the flow and pressure at the separator inlet.
Even with this limited scope, it is inevitable that every detail is not described in this specification. Additional information of general relevance may be added in future updates.

\subsection{Literature}

Offshore energy systems and solutions for \COO emission reduction from oil\&gas platforms is a topic that has been addressed and analysed in several studies in the literature \cite{KORPAS201218,NGUYEN2016673,NGUYEN2016325,riboldi2018,fard2018,10.3389/fenrg.2020.607284}, and several of these consider Norwegian offshore oil\&gas platforms. However, the cases and the assumptions are not documented in detail, making it difficult to test their conclusions and compare results. Lack of documentation may be due to confidentiality issues when working with real cases and data obtained from industry. Or it may be that even though the full data in principle is open, the studies only describe the subset of the data that is directly relevant for the discussion in the given study. 

Although we are not aware of any open datasets of a similar scope to what we present here, a lot of oil\&gas data is openly available, notably historical production data made available through the Norwegian Petroleum Directorate's data repository \cite{npddata}. 
%
Several open reservoir models \cite{lie_2019} that are relevant for analyses of subsurface dynamics also exist.
A general description of offshore processing systems and configurations can be found in \cite{Bothamley2004}. However, for exact modelling of power consumption, detailed information is necessary regarding produced fluid properties, export conditions and processing unit setpoints and performance. Still, such details are normally not available due to confidentiality. Some examples of modelled processing facilities and associated power consumption can be found in \cite{NGUYEN2016325,NGUYEN2013,NGUYEN2014,NGUYEN2014454,VOLDSUND201445}.  
%
Some electrical data and diagrams from real North Sea platforms have been published \cite{hadiya2011}, but again the description is incomplete and the underlying data are not provided.

\subsection{Contributions}

The main contribution of this article is
the specification of a realistic and open reference oil\&gas platform as a suitable study case that facilitates reproduction and comparison of results obtained with different methods and tools.
Additionally, actual implementations of this case in two different modelling environments are presented together with basic simulation results that show the platform in operation including some of its electrical characteristics. More thorough analyses are left for future publications.

\section{General specifications}
\label{sec:leogo}

\begin{figure*}
    \centering
    \includegraphics[width=15cm]{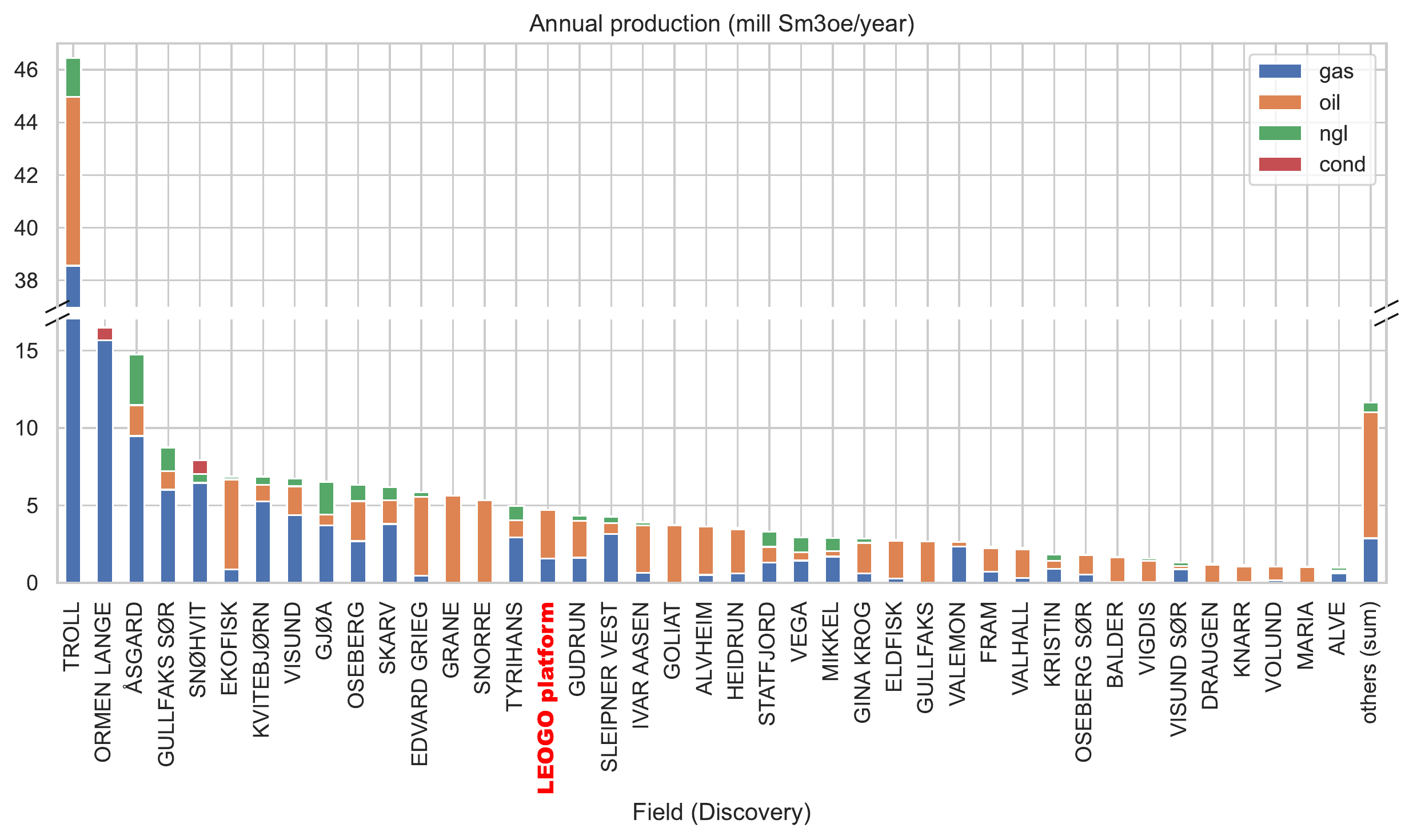}
    \caption{LEOGO reference platform compared to NCS platforms in 2018 in terms of production of natural gas, oil, natural gas liquids (ngl) and condensate (cond). (data source: Norwegian Petroleum Directorate)}
    \label{fig:leogo_vs_others}
\end{figure*}

\begin{figure*}
    \centering
    \includegraphics[width=15cm]{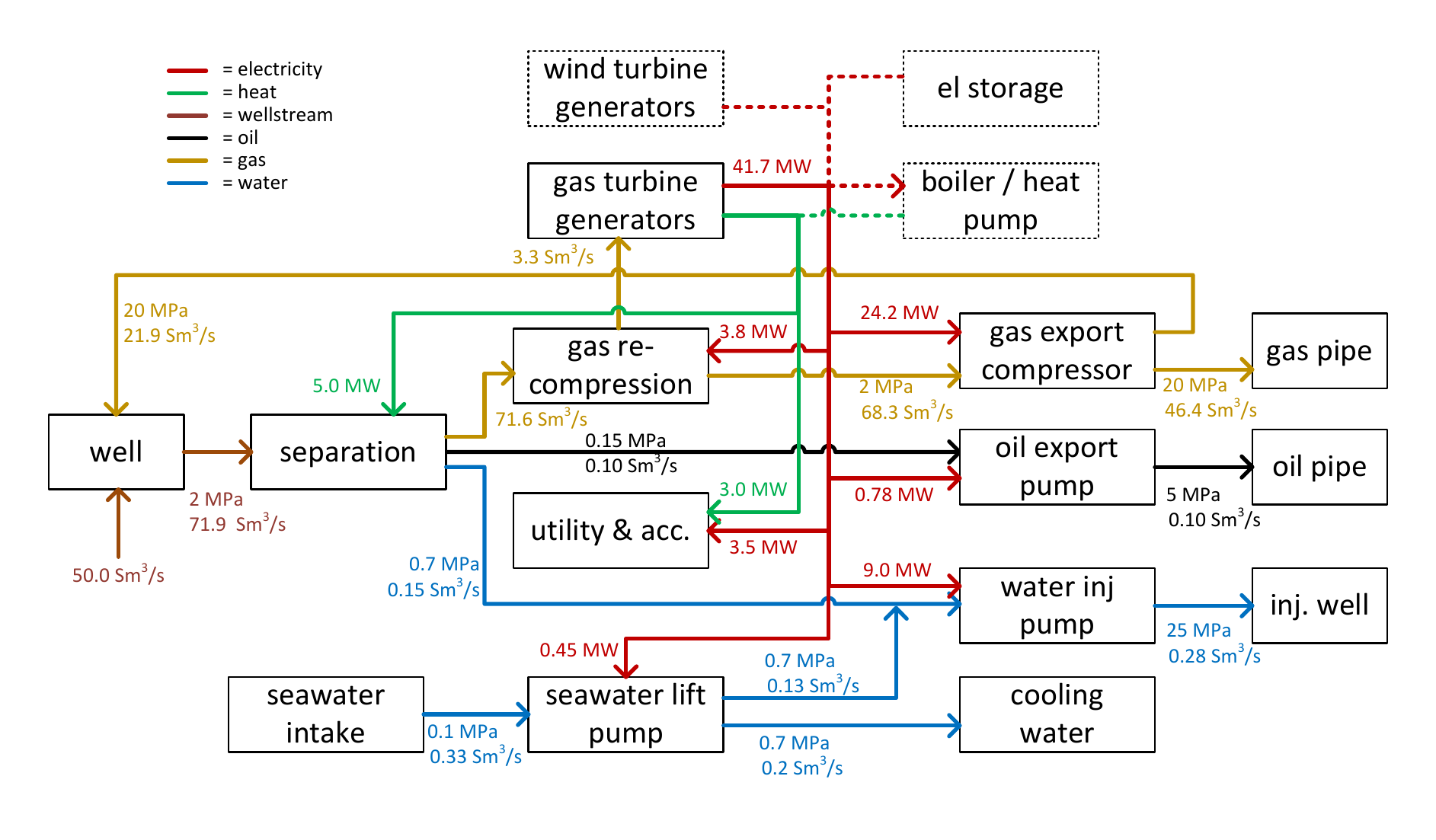}
    \caption{LEOGO reference platform: main components and energy/matter flows and pressure levels.}
    \label{fig:leogo}
\end{figure*}

The LEOGO reference platform is meant to be representative of what could be considered a typical oil\&gas platform on the Norwegian Continental Shelf~(NCS), with oil\&gas production rates compared to other fields as indicated in Figure~\ref{fig:leogo_vs_others}. 
Although a hypothetical field and platform,  its characteristics have been chosen to be \emph{realistic}.

The specification has a strong emphasis on the energy system, and what is relevant for energy system operational planning, power management and electrical stability, considering timescales of a month down to milliseconds. Reservoir dynamics and long-term changes in production rates and resulting changes in energy demand are not included.  
The data are available from a public repository \cite{leogodata}, and this specification may be updated and expanded in the future as needed.
For example, to make the case relevant for lifetime analyses and investment optimisation, energy demand variations over many years are needed.

A schematic of the platform's processing and energy system is shown in Figure~\ref{fig:leogo}, with details described in the following.

\subsection{Oil field and processing characteristics}

The LEOGO case represents an oil\&gas field with the main characteristics as shown in Table~\ref{tab:leogo_field}. 
The platform processes petroleum from multiple production wells, and
wellstream transport from the reservoirs up to the platform is assisted by gas lift.
Gas lift is used instead of electric pumps since it is more common on Norwegian fields. The wellstream pressure at the separator inlet is \unit[2]{MPa}, as indicated in Figure~\ref{fig:leogo}.
Reservoir pressure is maintained via water injection, using both water separated out from the wellstream, and additional seawater.

\begin{table*}[h]
\centering
\caption{Field and process characteristics}
\label{tab:leogo_field}
\begin{tabular}{lrl}
\hline
Description                  & Value   & Unit \\ \hline
Production rate, oil &          8600 & \si{Sm^3/day} \\
Production rate, gas &          $4.3\cdot 10^{6}$ & \si{Sm^3/day} \\
Production rate, water &        13000 & \si{Sm^3/day} \\
Separator inlet pressure & 2 & \si{MPa} \\ 
Separation stages       &   3   & \\
Gas pressure at export point & 20 & \si{MPa}  \\
Oil pressure at export point & 3 & \si{MPa}  \\
Wellstream gas to oil ratio (GOR) & 500 & \\
Wellstream water cut (WC) &         0.6 & \\
Natural gas \COO content &          2.34 & \si{kgCO_2/Sm^3} \cite{sandmo2016} \\
Natural gas energy content &        40 & \si{MJ/Sm^3} \\ \hline
\end{tabular}
\end{table*}

All compressors on the LEOGO platform have electric drives, which is usually the case for modern platforms. On many real older platforms, however, compressors are driven directly by gas turbines, making it more complicated to switch to another energy source than natural gas. The choice of having only electric drives is motivated by the fact that this allows more easily for a shift of the primary energy source from fossil fuel to renewables, which is the underlying motivation for this work. So either the LEOGO platform may be viewed as a relatively modern installation, or as an older one where the directly-driven compressors have been replaced by electrically driven ones (even though this is an expensive modification).

Separation and gas re-compression occur in three stages. These are integrated processes although they are illustrated with separate blocks in Figure~\ref{fig:leogo} for simplicity.

%

%

%

Electricity demand on the platform is dominated by gas compressors and water injection pumps. Note that some of the compressed gas is re-circulated into the well and used for gas lift.
The electricity and heat demand for utility and accommodation can be considered as independent from the production rates. The main heat demand is otherwise in the separation process.
Heat is provided via waste heat recovery from the gas turbines.
Flow rates and pressure levels are specified in Figure~\ref{fig:leogo}. All flow rates are given as volumetric flow at standard conditions (\unitfrac{Sm$^3$}{s}), where standard conditions are \unit[15]{$^\circ$C} and normal atmospheric pressure, \unit[101.325]{kPa}.

The characteristics for oil, natural gas and water fluids are shown in Table~\ref{tab:leogo_fluids_characteristics}.
Gas gravity, compressibility and temperature are needed to compute pressure drop in gas pipeline using the Weymouth formula.
Specific heat ratio and the individual gas constant are relevant for computing the energy demand by compressors.
The Darcy friction factors are used to compute pressure drop in pipelines when using the Darcy-Weissbach formula. In reality this factor is not a constant, but dependent on other factors (such as viscosity).

\begin{table*}
\centering
\caption{Fluid characteristics}
\label{tab:leogo_fluids_characteristics}
\begin{tabular}{lrl}
\hline
Description & value & unit \\
\hline
Gas compressibility & 0.9 & \\
Gas energy value  & 40 & \si{MJ/Sm^3} \\
Gas heat capacity ratio & 1.27 & \\
Gas individual gas constant & 500 & \si{J/kg K} \\
Gas gravity         & 0.6 & \\
Gas density & 0.84 & \si{kg/Sm^3} \\
\hline
Oil density & 900 & \si{kg/m^3} \\
Oil viscosity & 0.0026 & \si{\kilogram/\meter\second} \\
Oil darcy friction & 0.02 & \\
\hline
Water density & 1000 & \si{kg/m^3} \\
Water darcy friction & 0.01 & \\
\hline

\end{tabular}
\end{table*}

\subsection{Energy supply}

Regarding energy supply, multiple alternatives are defined: The \emph{base case} represents a traditional system with energy supply from gas turbine generators and a set of \emph{variations} represent systems with modified electric power supply. Presently, two such variations are defined, as shown in Table~\ref{tab:basecase_variations}.

\begin{table*}[h]
    \centering
    \caption{Electric power supply variations}
    \label{tab:basecase_variations}
    \begin{tabular}{llrl}
        \hline
        Case        &  \multicolumn{2}{l}{Power}   & Type\\
        \hline
        Base case   & 3x    & \unit[21.8]{MW} & gas turbines\\
        \hline
        Variation~A & 3x    & \unit[21.8]{MW} & gas turbines\\
                    & 3x    & \unit[8.0]{MW} & wind turbines\\
        \hline
        Variation~B & 3x    & \unit[21.8]{MW} & gas turbines\\
                    & 3x    & \unit[8.0]{MW} & wind turbines \\
                    & 1x    & \unit[4.0]{MW} & battery (\unit[4]{MWh}) \\
        \hline
    \end{tabular}
\end{table*}

\begin{figure*}
    \centering
    \includegraphics[width=15cm]{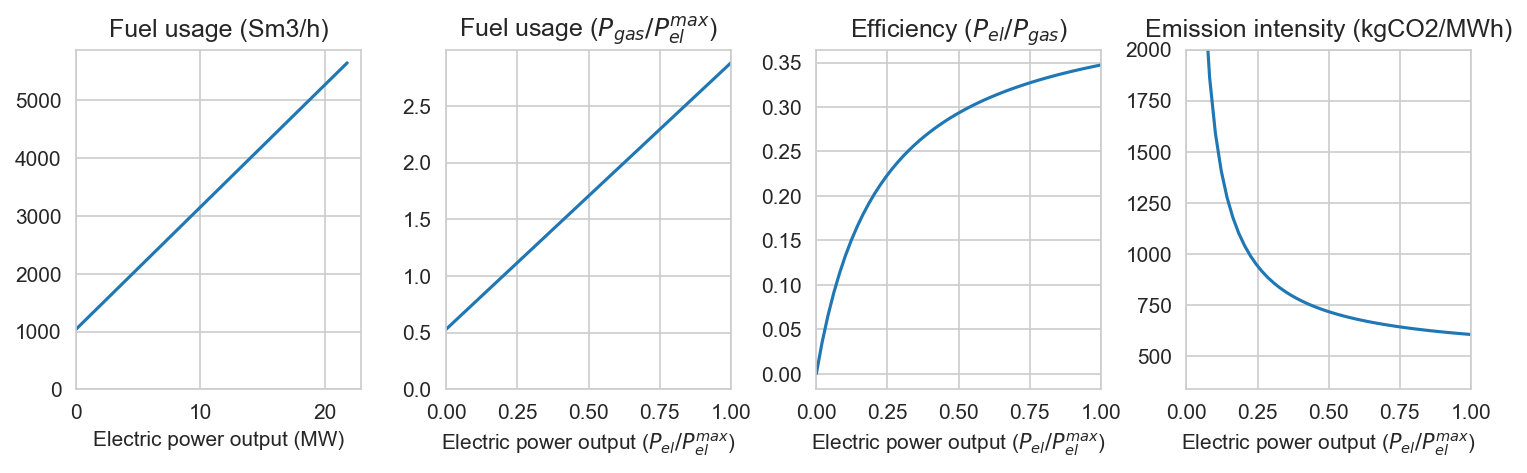}
    \caption{Gas turbine generator fuel usage, efficiency and emission intensity versus power output, $P_\text{el}$. $P_\text{gas}$ is the energy content of the gas fuel, and $P_\text{el}^\text{max}$ is the generator capacity.}
    \label{fig:gtg_efficiency}
\end{figure*}

The gas turbine generators resemble GE LM2500 gas turbine generators with rated power of \unit[28]{MVA} (\unit[21.8]{MW} active power), ramping rates of \unit[100]{\%} per minute, and minimum loading of \unit[3.5]{MW}.
Startup time is set to be 15 minutes preparation plus 15 minutes synchronisation time, so a total of 30 minutes from activation to delivery of electric power \cite{GONZALEZSALAZAR20181497,gasturbineswartsila}.
Gas turbine fuel consumption is linearly related to its electric power output, with efficiency \unit[34.7]{\%} at full load and \unit[20]{\%} at \unit[20]{\%} loading \cite{HAGLIND20091484}, see Figure~\ref{fig:gtg_efficiency}.

\unit[8]{MW} wind turbines are used for the case variations. And for Variation~B with a battery storage system, energy and power capacities of \unit[4]{MWh}/\unit[4]{MW} are used, with a round-trip efficiency of \unit[90]{\%}.

To have some operational margin and be able to cope with unforeseen variations in power demand, the system is operated with a reserve requirement of \unit[5]{MW}. This is unused online power capacity that can be activated automatically to maintain power balance.

\subsection{Operating state}

The electricity supply and demand by different main components in nominal operation in the base case is shown in Table~\ref{tab:demand}. 
The difference between load and generation are mainly electrical losses of the cables, transformers and converters. 
A small part of the deviation stems from the voltage-dependence of the loads, which means that the loads slightly change their power consumption when the real voltage is not exactly the rated voltage.

Fluid and energy flow rates are indicated in Figure~\ref{fig:leogo}.
The relationship between flow rates, pressure levels and power demand by pumps and compressors has been computed assuming 
water pump efficiency of 0.75, 
oil pump efficiency of 0.6 and 
compressor efficiency of 0.75. 
These relationships are elaborated on in Section~\ref{sec:pump_compressor_demand}.
With the high loading of gas compressors, the operating state may be considered to represent the early plateau production phase with a high production rate.

In this operating situation, two gas turbines could be just enough to cover the electricity demand, but the \unit[5]{MW} reserve requirement means that all three gas turbines are required to be online in normal operation.
The total heat demand is \unit[8]{MW}, and is covered by waste heat recovery from the gas turbines, see Table~\ref{tab:heat}.

\begin{table*}
\centering
\caption{Electric power capacities and base case loading.}
\label{tab:demand}
\begin{tabular}{lSS}
\hline
Description                  & {Capacity (MW)}   & {Loading (MW)} \\ \hline
Sea water lift pump 1 (SWL1) & 0.75     & 0.45  \\  
Sea water lift pump 2 (SWL2) & 0.75     & 0 \text{(off)} \\ 
Sea water lift pump 3 (SWL3) & 0.75     & 0 \text{(off)}  \\
Air compressor 1 (ACO1)     &   1.3     & 0.5 \\ 
Air compressor 2 (ACO2)     &   1.3     & 0.5 \\ \hdashline
Gas export compressor 1 (GEX1)  & 8.2  & 8.07  \\ 
Gas export compressor 2 (GEX2)  & 8.2  & 8.07  \\
Gas export compressor 3 (GEX3)  & 8.2  & 8.07  \\
Oil export pump 1 (OEX1)    & 1.5   & 0.39  \\
Oil export pump 2 (OEX2)    & 1.5   & 0.39  \\
Water injection pump 1 (WIN1)   & 4.8   & 3.0 \\ 
Water injection pump 2 (WIN2)   & 4.8   & 3.0 \\ 
Water injection pump 3 (WIN3)   & 4.8   & 3.0 \\ 

Gas re-compressor 1 (REC1)  & 1.5  & 1.27 \\ 
Gas re-compressor 2 (REC2)  & 1.5  & 1.27 \\
Gas re-compressor 3 (REC3)  & 1.5  & 1.27 \\  \hdashline

Utility ASM \unit[690]{V} A (ASM1) & 0.25    & 0.2   \\ 
Utility ASM \unit[690]{V} B (ASM2) & 0.25    & 0.2   \\

Utility load \unit[690]{V} A (LOD1) & 2.75   & 1.05   \\ 
Utility load \unit[690]{V} B (LOD2) & 2.75   & 1.05   \\
Utility load \unit[400]{V} A (LOD3) & 0.5    & 0.25 \\ 
Utility load \unit[400]{V} B (LOD4) & 0.5    & 0.25 \\ \hdashline
Drill 1 (DRL1) & 0.8 & 0 \text{(off)} \\ 
Drill 2 (DRL2) & 0.8 & 0 \text{(off)} \\ 
Drill 3 (DRL3) & 0.8 & 0 \text{(off)} \\ 
Drill 4 (DRL4) & 0.8 & 0 \text{(off)} \\ 
Drill 5 (DRL5) & 0.8 & 0 \text{(off)} \\ 
Drill 6 (DRL6) & 0.8 & 0 \text{(off)} \\  \hdashline
Consumption deviation  &   & 0.12  \\
Converter, transformer and line losses &   & 0.81 \\

\hline
SUM electric power demand   &           & 43.18   \\   
\hline
Gas turbine 1           & 21.8    & 14.39  \\ 
Gas turbine 2           & 21.8    & 14.39  \\
Gas turbine 3           & 21.8    & 14.39  \\ \hdashline
Wind turbine 1           & 8.0    & 0 \text{(off)}  \\ 
Wind turbine 2           & 8.0    & 0 \text{(off)}  \\
Wind turbine 3           & 8.0    & 0 \text{(off)}  \\ \hline
SUM electric power supply &       & 43.18  \\ \hline
\end{tabular}
\end{table*}


\begin{table}
\centering
\caption{Heat supply and demand.}
\label{tab:heat}
\begin{tabular}{lS}
\hline
Description                       & {Loading (MW)} \\ \hline
Separation/processing             & 5.0 \\
Utility and accomodation          & 3.0 \\ \hline
SUM heat demand                   & 8.0 \\ 
\hline
Gas turbine 1                &  15.2  \\ 
Gas turbine 2                &  15.2 \\
Gas turbine 3                &  15.2 \\ \hline
SUM heat supply                  &  45.6 \\ \hline
\end{tabular}
\end{table}

\subsection{Compressor and pump power demand}
\label{sec:pump_compressor_demand}

Pump power demand $P_\text{pump}$ can be computed according to the relationship
\begin{equation}
    P_\text{pump}=\frac{1}{\eta}q\Delta p,
\end{equation}
where 
$\eta$ is efficiency, 
$q$ is flow rate and 
$\Delta p=p_2-p_1$ is inlet/outlet pressure difference.
For the water injection pumps with 
$\Delta p=(25-0.7)$~MPa, 
$q=\SI{0.277}{Sm^3/s}$, and
$\eta=0.75$ 
this gives $P_\text{pump}=8.97$~MW.
Similarly, for the oil export pumps with 
$\Delta p=(5-0.3)$~MPa, 
$q=\SI{0.098}{Sm^3/s}$, and 
$\eta=0.6$ 
it gives $P_\text{pump}=0.79$~MW.

Gas compressor power demand $P_\text{comp}$ is computed assuming an adiabatic process and ideal gas, with the equation
\begin{equation}
    P_\text{comp} = \frac{1}{\eta}\frac{\rho ZRT}{k-1} q \Bigl[ (\frac{p_2}{p_1})^a-1\Bigl],
\end{equation}
where
$\rho=0.84$ is the natural gas density,
$Z=0.9$ is the gas compressibility,
$R=$\unit[500]{J/kg K} is the individual gas constant,
$T=300$~K is the inlet temperature,
$k=1.27$ is the gas specific heat ratio, and 
$a=\frac{k-1}{k}$.
For the main gas export compressors with $\eta=0.75$, $p_1=$\unit[2]{MPa}, $p_2=$\unit[20]{MPa}, $q=\unit[68.3]{Sm^3/s}$ this gives $P_\text{comp}=$\unit[24.2]{MW}

For gas re-compression in the separator train, we determine the power demand by assuming a single separation stage and intermediate gas pressure of $p_1=\unit[1.3]{MPa}$. With an outlet pressure of $p_2=\unit[2]{MPa}$ we find using the same formula a power consumption of $P_\text{comp}=\unit[3.8]{MW}$.

Note that several simplifications are made here: 
In a real system, usually the separation train consists of several separators at different pressure levels, and so does the re-compression train. Consequently, the gas rates leaving each separator stage and following gas rates per compressor stage will be different. Most free gas, including lift gas which mainly consists of light components, leaves the separation train in the first stage at higher pressure. The pressure for the oil, however, will be reduced to almost atmospheric conditions in order to remove remaining gas. Efficiencies for pump and compressor systems will also depend on system configuration and ability to control these units. These assumptions will lead to some deviation from a real system, which should be acceptable at this point where the focus is on the electrical part.

\subsection{Time-series data}
\label{sec:leogo_timeseries}

The variability in total wellstream flow rate (production rate) and hence energy demand as well as the variability in wind power availability are given by time-series data.
Data provided with the LEOGO specification \cite{leogodata} includes time-series with 1-month duration and 1 minute resolution.

For flow rate / energy demand, the wellstream flow variation is set such that the power demand by compressors and pumps, following a linear dependency on the flow rate, varies $\pm 4$~\% with a period of about 25 minutes. 
This is based on inspection of power data from a real case platform. Such power variations are not a generic characteristic, but highly dependent on the specific case and the type of activities and equipment in use. They are nevertheless included in the present dataset to have some reasonable flow and power demand variation without being concerned about their origin.
We consider this acceptable as studies based on this platform specification are likely to be mainly concerned with the energy system and not the processing system.
If the processing system is not represented in the model, which is the case for an electric only model, the implementation of the variability is that the energy demand in pumps and compressors vary according to the same time-series.

Wind power data has been obtained from publicly available met mast measurements from Sulafjorden at the coast of Norway
\cite{winddata}.
Measurements used are wind speeds at height 92.5 m with a resolution of 10 measurements per second (\unit[10]{Hz} sample rate). 
A coastal fjord location is not fully representative of offshore conditions, with higher levels of turbulence and variation, but is used for lack of better high-resolution wind data that is publicly available and relevant for North Sea locations.

Wind forecasts with a 30 minute resolution have been created by re-sampling the wind speed measurements and then adding random noise to represent forecast error.
This results in a root mean square error (RMSE) deviation of 3.3 m/s between forecast and actual wind speeds at a resolution of 1 minute.
Wind power has been computed using the power curve of a real 8 MW offshore wind turbine.
Wind speed measurement data and calculated wind power data for March and April 2020, down-sampled to 1 minute resolution are included with the LEOGO dataset and used in the simulations reported here.

\begin{figure*}
    \centering
    \includegraphics[scale=0.65]{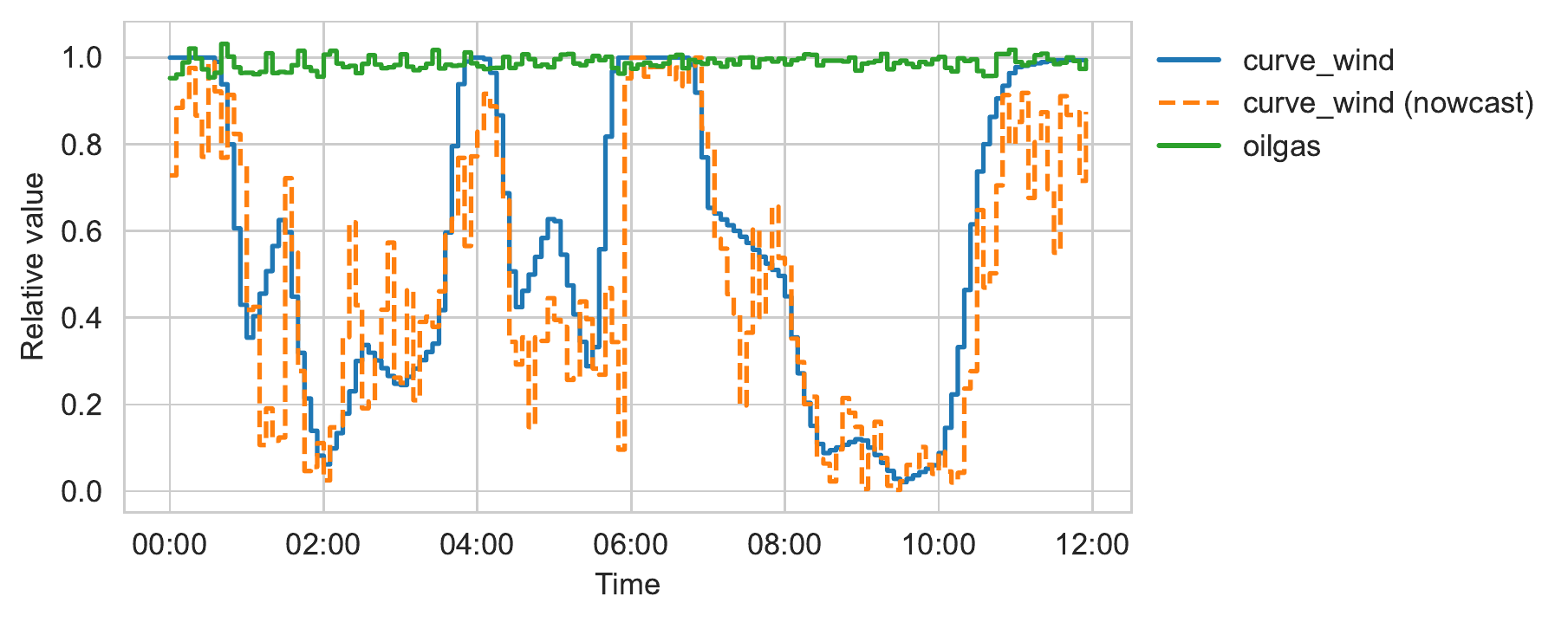}
    \caption{Normalised profiles. The figure shows a 12 hour extract of the data, with 5 min resolution.}
    \label{fig:profiles}
\end{figure*}

An extract of the time-series is shown in Figure~\ref{fig:profiles}. The profiles could be considered as forecasts used for planning. In actual operation, there will be some deviations from these forecasts. The \emph{nowcast} profile for wind power should be interpreted as updated closer to real-time forecast with reduced forecast error.

\section{Electrical system}
\label{sec:electrical_system_model}

\begin{figure*}[h]
    \centering
    \includegraphics[width=15cm]{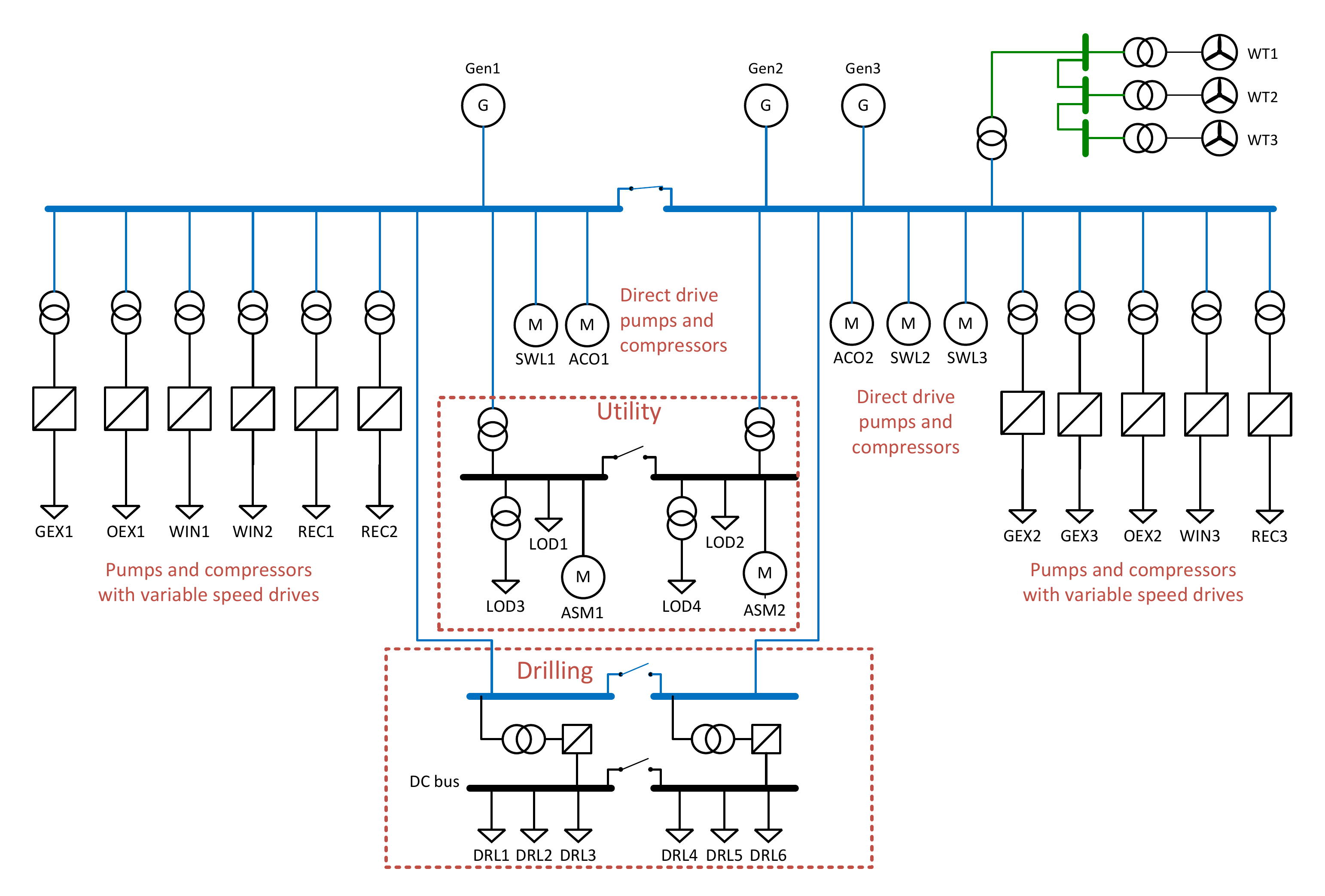}
    \caption{Electrical model}
    \label{fig:electricalmodel}
\end{figure*}

A full description of the electrical system requires a lot of detail regarding the properties of cables and transformers making up the electrical grid and the properties of loads and generators connected to it. For analyses of dynamic behaviour, generator control systems play an important role. The same is true for the control systems of other active components, such as variable speed drives for pumps and compressors.
In the following, an overview of the various components in the electrical system is given. The system has been modelled in DIgSILENT PowerFactory and this freely available model \cite{leogodata} includes further details not given here. 
A simplified single-line diagram of the electrical model is presented in Figure~\ref{fig:electricalmodel}.

\subsection{Voltage levels}
\label{sec:VoltageLevels}

The platform has four AC voltage levels and one DC voltage level:
\begin{itemize}
    \item \SI{33}{kV_{AC}} is the voltage that is used for the collection grid of the wind turbines and the transmission to the platform.
    \item \SI{11}{kV_{AC}} is the voltage at the main busbar system. The gas turbine generators and several of the large loads are connected either directly or via dedicated transformers to this busbar system.
    \item \SI{690}{V_{AC}} is used for the utility busbar system is where smaller pumps and compressors, and auxiliary equipment is connected. It is fed from the \unit[11]{kV} system via transformers.
    \item \SI{400}{V_{AC}} is the common low voltage level, and it is used here mainly for the accommodation facilities. The \unit[400]{V} system is fed from the \unit[690]{V} system via transformers.
    \item \SI{+1300}{V_{DC}} is used in addition to the AC voltge levels, where a drilling switchboard is operated at this voltage level. This DC busbar system is fed through active power converters in rectifier mode, that connect via the converter transformers to the \unit[11]{kV} system. The DC system is monopolar, and the return path is through earth.
\end{itemize}

The \unit[11]{kV}, \unit[690]{V} and \SI{+1300}{V_{DC}} busbar systems are realised as split single busbars with tie breaker. The \SI{33}{kV_{AC}} is realised with simple single busbars.
The \unit[400]{V} level consists of two separate single busbars without a tie breaker.
All tie breakers are normally open, besides the \unit[11]{kV} main busbar system which is normally closed.

\subsection{Transformers}

There are four \emph{stand-alone} transformers on the platform. 
In this context stand-alone means that they directly connect two busbar systems of different voltage, in contrast to the other converter transformers that are dedicated to connecting a power electronic converter.
\begin{itemize}
    \item The two larger units supply electricity from the \unit[11]{kV} main busbar system to the \unit[690]{V} utility busbar system. These transformers have a rating of \unit[3.3]{MVA} each, a short circuit voltage of \unit[11]{\%}, with \unit[0.35]{\%} losses. They have delta windings on the high voltage side, and star windings at the low voltage side, with the star point earthed with a \SI{4}{\ohm} resistor (\SI{100}{A})
    \item The two smaller units supply the accommodation facilities at \unit[400]{V} from the \unit[690]{V} utility busbar system. These transformers have a rating of \unit[600]{kVA} each, a short circuit voltage of \unit[6]{\%}, with \unit[1]{\%} losses. They have a similar delta-star configuration, but their star point is directly earthed.
\end{itemize}

\subsection{Cables}

Three AC cable types and one DC cable type are used on the platform:

\begin{itemize}
    \item \unit[33]{kV} \unit[3x240]{\smm} XLPE-insulation copper-conductor AC cables are used for the wind turbines. This cable type has a rated current of \unit[489]{A}.    
    \item \unit[11]{kV} \unit[3x120]{\smm} XLPE-insulation copper-conductor AC cables are used for the majority of applications. This cable type has a rated current of \unit[360]{A}. 
    \item \unit[11]{kV} \unit[3x240]{\smm} XLPE-insulation copper-conductor AC cables are used where a higher current rating of \unit[520]{A} is needed. For the connection of each gas turbine generator, four parallel cables of this type are used. This cable type is (for simplicity) also used for the two short low voltage connections between the \unit[690]{V} busbar system and the \unit[690-400]{V} transformer at the accommodation facilities (even though a lower voltage rating would be sufficient there).
    \item \SI{1.5}{kV_{DC}} \unit[1x500]{\smm} XLPE-insulation copper-conductor DC cables are used for the connection between the drilling DC busbar system and the rectifiers that feed power to it. Four cables in parallel are used for each connection, due to the high currents involved.
\end{itemize}

\subsection{Gas turbine generators}

The platform has three gas turbine generators with a rated power of \unit[28]{MVA}, rated voltage of \unit[11]{kV} and rated power factor of 0.85. 
They have star windings, and the star point is earthed with \SI{127}{\ohm} (\SI{50}{A}).
The generators are directly connected to the main busbar system. 
The gas turbines are controlled by governors and the synchronous generators by static excitation systems. 
All three turbine-generator-transformer setups are identical.

\subsection{Wind turbine generators}
 
The gas-turbine-based electricity supply of the LEOGO platform is accompanied by three \SI{8}{MW} direct drive type-4 wind turbines. The rated power factor is 0.9, and the generator voltage is \unit[690]{V}, which is transformed to \unit[33]{kV} by a dedicated transformer.
The wind turbine model attempts to resemble the Siemens Gamesa SG 8.0-167 DD wind turbine, as this turbine type is used for the same purpose on a real offshore installation (the Hywind Tampen project \cite{hywindtampen}). 
However, as no model of the SG 8.0-167 DD is openly available, the similarity is limited to the general parameters like power rating, turbine type, voltage level, power curve, e.t.c. 
The details of the implemented wind turbine model just resemble a generic wind turbine.

\subsection{Induction motors}

There are three types of directly grid-connected induction motors:
\begin{itemize}
    \item Three sea water lift pumps (SWL) are driven by \unit[750]{kW} \unit[11]{kV} induction motors with a rated power factor of 0.87 and an efficiency of \unit[96.9]{\%}
    \item Two air compressors (ACO) are driven by \unit[1.3]{MW} \unit[11]{kV} induction motors with a rated power factor of 0.9 and an efficiency of \unit[95.9]{\%}
    \item Two induction motors (ASM) at \unit[0.25]{MW} and \unit[690]{V} with a rated power factor of 0.92 and an efficiency of \unit[96.1]{\%}, generally representing one of the low voltage machines at the utility busbars.  
\end{itemize}

All induction motors that are a part of a variable speed drive are not modelled in detail, as explained in \cref{sec:VSDs}.

\subsection{Variable speed drives}
\label{sec:VSDs}

The \emph{regular} (with dedicated power supply) variable speed drives consist of induction motors that are interfaced with a transformer and a back-to back AC-DC-AC converter.
The transformer is not earthed and draws electric power from the main \unit[11]{kV} busbar and reduces the voltage to a suitable level for the drive. 
This reduced AC voltage is then rectified by the active grid-side converter to provide the AC-DC-AC converters internal DC voltage. 
The machine-side converter is supplied from this internal DC voltage, and it drives the machine with variable frequency and variable voltage AC. 

There are three types of variable speed drives:

\begin{itemize}
    \item \unit[8.2]{MW} is the rating of the largest VSDs. Their transformer is delivering \unit[3.3]{kV} from the main \unit[11]{kV} busbar, and this is then rectified to \SI{+6}{kV_{DC}} by the active grid side converter
    \item \unit[4.8]{MW} is the rating of the medium-size VSDs. They operate at the same voltage levels as the \unit[8.2]{MW} units, just at lower currents.
    \item \unit[1.5]{MW} is the rating of the smaller VSDs. Their
    transformer delivers \SI{690}{V_{AC}} from the main \unit[11]{kV} busbar. This \SI{690}{V_{AC}} is then rectified to \SI{+1300}{V_{DC}} by the active grid side converter.
\end{itemize}

In addition to these \emph{regular} VSDs, there are the drilling VSDs that share a common power supply. 
There are two parallel transformer-converter systems where the \unit[3.3]{MVA} transformer (not earthed) draws power from the \unit[11]{kV} busbar, transforms it to \unit[690]{V}, and the active power converter rectifies it, and supplies the \SI{+1300}{V_{DC}} busbar system, as mentioned in \cref{sec:VoltageLevels}. 
This common DC busbar is different as compared to the other VSDs, where the DC voltage is only internal to each back-to-back converter. 
The machine side converters of the drilling VSDs are supplied from the common DC busbar, and they drive the induction machines.

For all VSDs, the induction machine and the machine-side converter are not yet modelled in detail; they only appear as an ideal DC loads, which is how they behave as seen from the grid in steady state operation.
This simplification was judged acceptable, as the back-to-back converter setup of the VSDs sufficiently decouples the grid dynamics from the rotating machine.
It is, however, planned to improve the model in the future by adding these details.

\subsection{General low voltage loads}

All the smaller loads (less than \unit[250]{kVA}) are not modelled one by one in detail, but represented as aggregated loads. There are two types of these aggregated loads:

\begin{itemize}
    \item Two \SI{690}{V_{AC}} loads at the utility busbars with a rating of \unit[2.75]{MW} and an actual consumption of \unit[1.05]{MW}, and a power factor of 0.9. \unit[60]{\%} of this load is induction motors, while the remaining \unit[40]{\%} are static frequency-independent loads with the composition \unit[28]{\%} constant power, \unit[4]{\%} constant current and \unit[8]{\%} constant resistance.
    \item Two \SI{400}{V_{AC}} loads at the accommodation busbars with a rating of \unit[0.5]{MW} and an actual consumption of \unit[0.25]{MW}, and a power factor of 0.98. \unit[10]{\%} of this load is induction motors, while the remaining \unit[90]{\%} are static frequency-independent loads with the composition \unit[36]{\%} constant power, \unit[9]{\%} constant current and \unit[45]{\%} constant resistance.
\end{itemize}

\section{Examples of simulation results}

Simulation results from both the operational optimisation model and the electrical model are presented in this section. The purpose of this is not the results in itself, but to demonstrate that the LEOGO platform specification makes sense and delivers reasonable results.

\subsection{Operational optimisation model}

\begin{figure*}
    \centering
    \includegraphics[width=15cm]{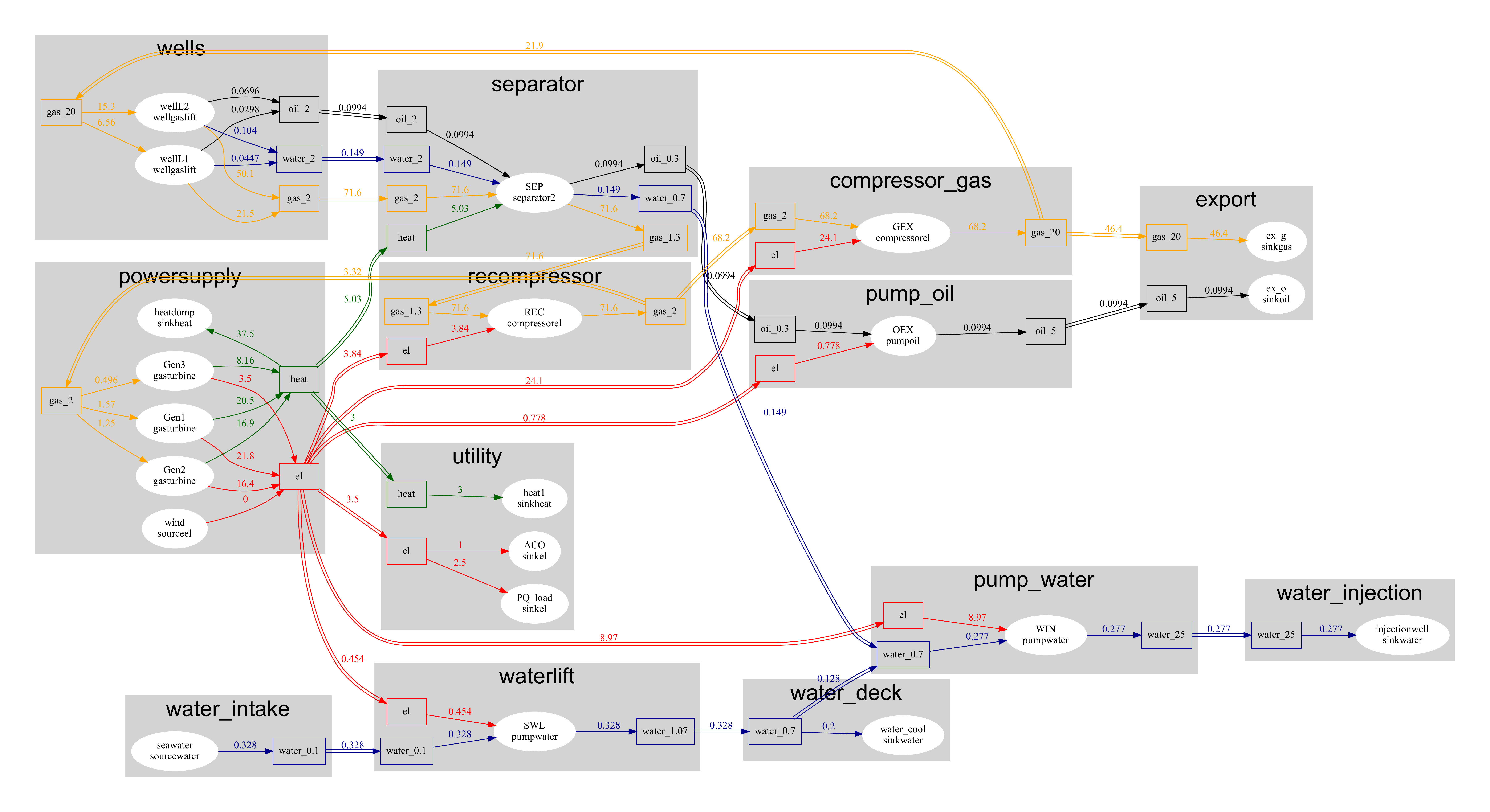}
    \caption{Overview of the LEOGO model represented in Oogeso. Edge colours represent different energy carriers, and numbers indicate flow ($\unit{Sm^3/s}$ or \unit{MW}). The numbers in the square boxes indicate inlet/outlet pressure levels (\unit{MPa}).}
    \label{fig:diagram}
\end{figure*}

The platform specification above has been implemented as an input dataset \cite{leogodata} for analyses with the openly available Oogeso tool \cite{oogeso}.
The system as represented in Oogeso is illustrated in Figure~\ref{fig:diagram}. As the figure shows, the emphasis is on the energy system, with a rather simplified description of the processing. For example the separation process is represented by a single unit, and the multi-phase wellstream flow is represented by its oil, gas and water components. 

The main reason for including the processing system is to capture the link between energy demand and oil\&gas production, and its main constraints. This will be relevant when considering the potential for demand-side flexibility to allow higher share of wind energy supply with minimum energy storage requirements.


The Oogeso tool applies a rolling-horizon mixed-integer linear optimisation model to schedule generation output and start and stop signals based on forecasts of demand and energy availability. The rolling horizon is to account for delays in ramping and flexibility in storage utilisation or demand. 

To demonstrate that the LEOGO specification makes sense, two simple simulations have been run over a timespan of 1000 minutes with timesteps of 5 min. The first is a modified version of the base case (see Table~\ref{tab:basecase_variations}) where the wellstream flow is reduced by \unit[20]{\%} between the 500 and 750 minute points. The second simulation represents Variation~A and includes wind turbines with their associated variability.

\begin{figure*}[h!]
    \centering
    \begin{subfigure}[b]{15cm}
        \includegraphics[scale=0.65]{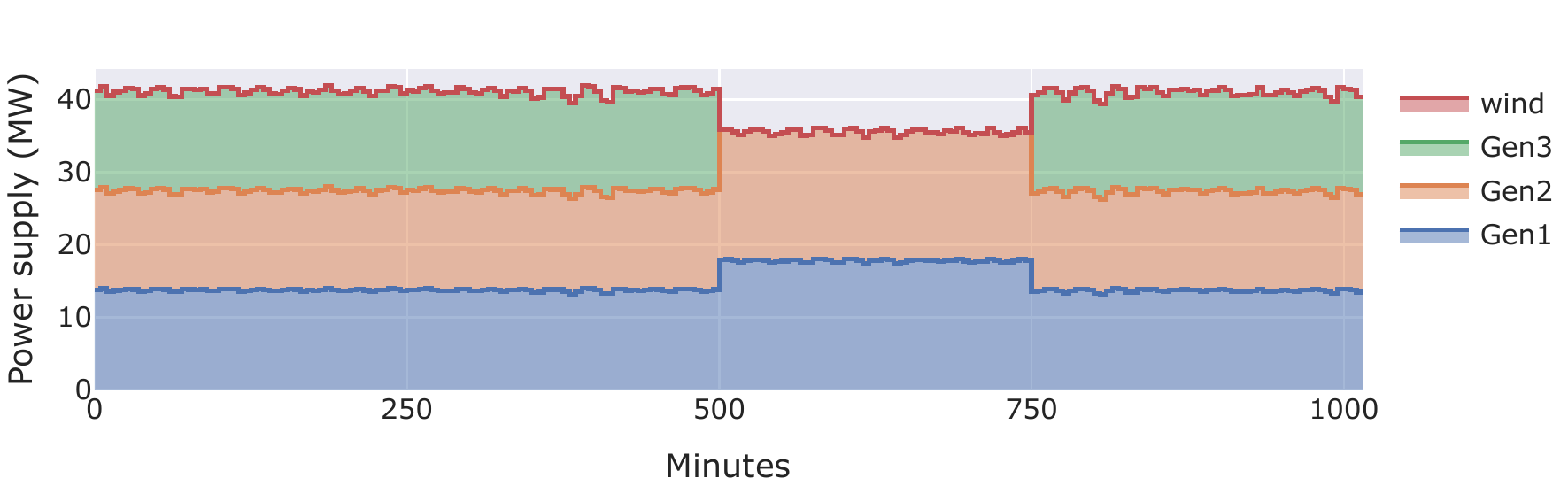}
        \caption{Modified base case}
    \end{subfigure}
    \begin{subfigure}[b]{15cm}
        \includegraphics[scale=0.65]{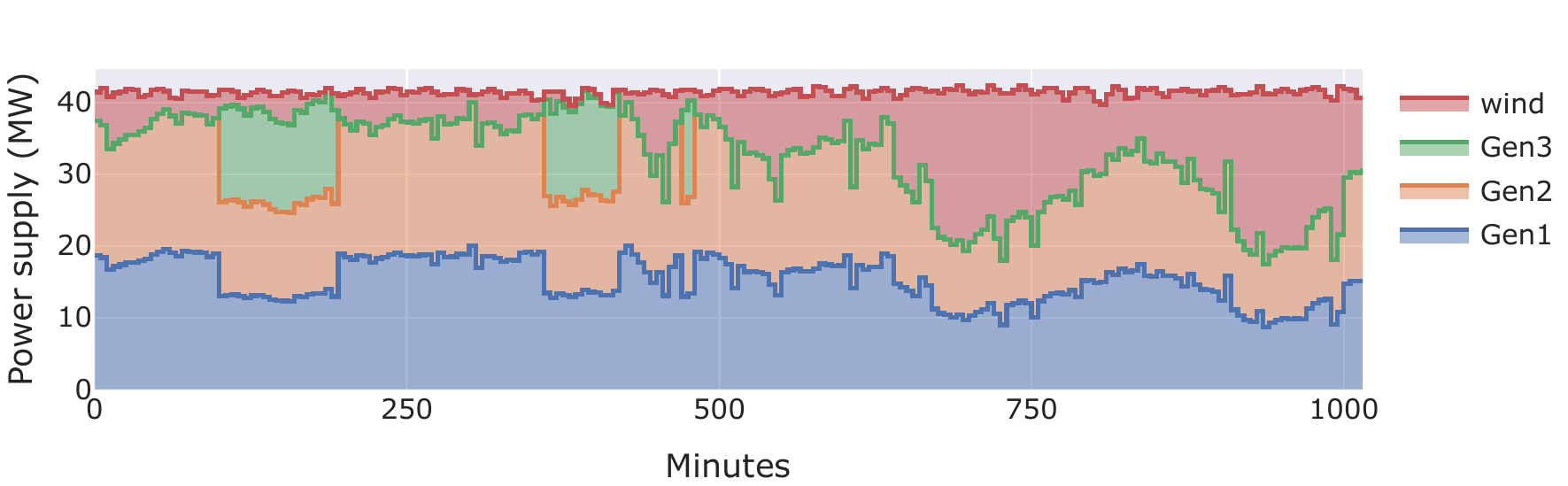}
        \caption{Wind case (Variation~A)}
    \end{subfigure}
    \caption{Power supply from each generator in the two simulation cases.}
    \label{fig:powersupply}
\end{figure*}

Figure~\ref{fig:powersupply} shows supply by source in the two simulation cases. The observed small fluctuations in total power demand and supply are due to the small variations in the input wellstream flow rate data.  
In the base case, we see that initially all three gas turbines are online, sharing the load evenly.
As the wellstream flow drops after 500 minutes, the energy demand in compressors and pumps also drops and then only two gas turbines are needed. Switching off the third generator is beneficial because the efficiency of the gas turbines is higher at higher loading, resulting in lower total fuel consumption and therefore lower \COO emissions per output electric power.
In the case with wind power, there is more variability and multiple start and stops of the third gas turbine.

\begin{figure*}[h]
    \centering
    \begin{subfigure}[b]{15cm}
        \includegraphics[scale=0.65]{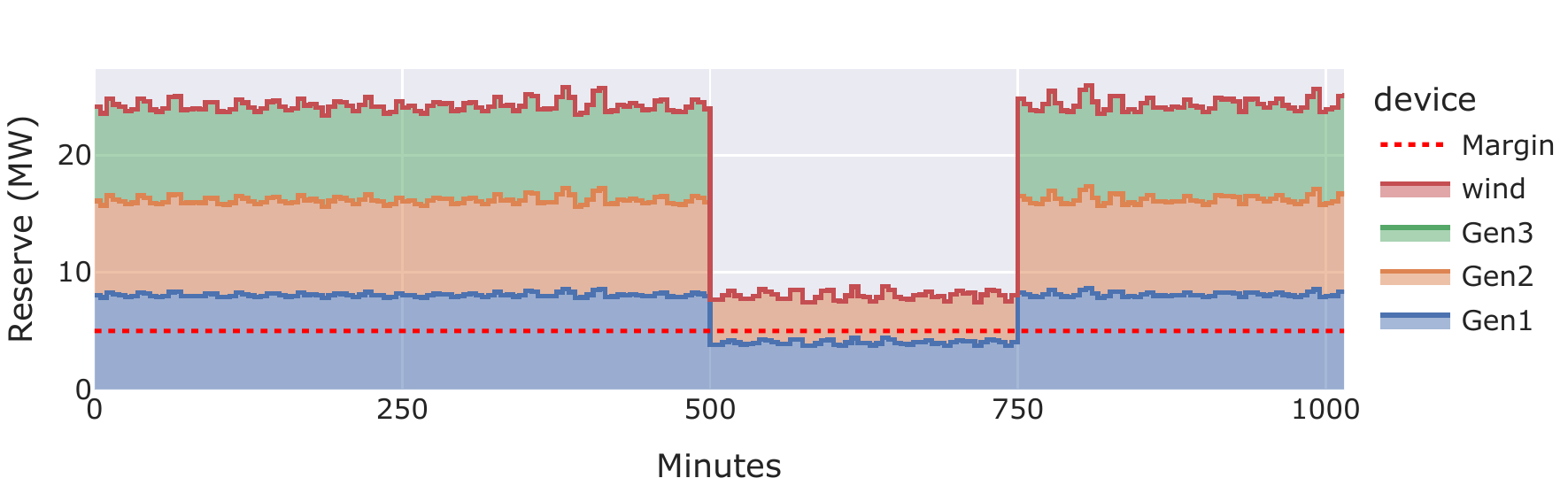}
        \caption{Modified base case}
    \end{subfigure}
    \begin{subfigure}[b]{15cm}
        \includegraphics[scale=0.65]{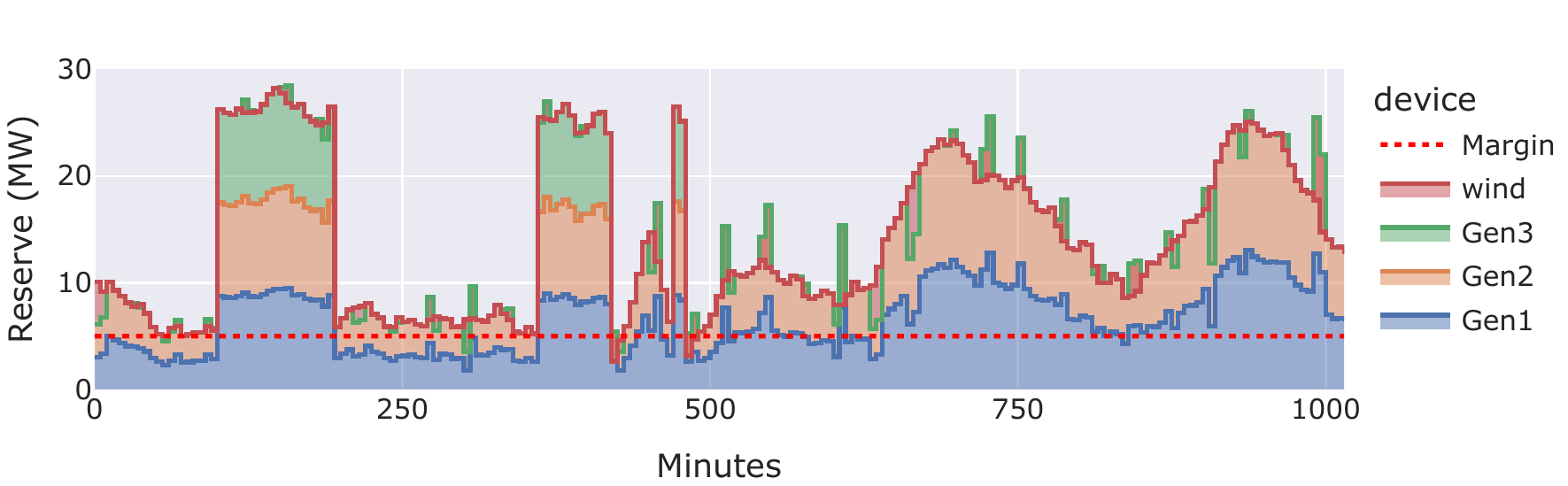}
        \caption{Wind case (Variation~A)}
    \end{subfigure}
    \caption{Online reserve provided by each generator in the two simulation cases.}
    \label{fig:reserve}
\end{figure*}

The available online power reserve is shown in Figure~\ref{fig:reserve}. In the base case, there is plenty of available reserve when all three gas turbines are running.
When the power demand drops after 500 minutes, the third gas turbine is switched off, as the two remaining gas turbines are then sufficient to both cover the load and provide the required reserve.
At 750 minutes, the power demand increases and then the third gas turbine is switched on again. Although two gas turbines could still provide sufficient power to meet the demand, there wouldn't be enough reserve without the third gas turbine.

In the case with wind power, we see a similar situation with plenty of available reserve when all three gas turbines are on, and operation close to the threshold otherwise.

A thorough analysis of different configurations and operating strategies of the LEOGO platform is planned for a separate publication.

%

\subsection{Electrical model}
\label{sec:electricalmodelimplementation}

The electrical system (more precisely, the electro-mechanical system), as discribed in Section~\ref{sec:electrical_system_model}, has been implemented in the DIgSILENT PowerFactory simulation tool.
As this model contains no proprietary information, it has been made public and can be freely downloaded \cite{leogodata}, used and modified.
This RMS domain model is suitable to assess electrical system behaviour in more detail, allowing to study power flow, electro-mechanical transients, outer-level controls, stability indices, etc. 

To display some of the functionalities of the model, the event of the loss (disconnection) of a gas turbine has been simulated for the base case (three gas turbines only) and the Variation~A (three gas turbines and three wind turbines). 
The battery of Variation~B has not yet been implemented in the electrical model.
In the base case, wind power output is zero, and all three gas turbines operate, of which one is lost. 
In Variation~A, the three wind turbines operate at full power output, and only two gas turbines operate, of which one is lost, and the other having to handle the incident alone.
It should be noted that the disconnection of a gas turbine is a critical event, that stresses the electrical power system beyond the capability of the \unit[5]{MW} operational reserve.

\begin{figure}[h]
    \centering
    \input{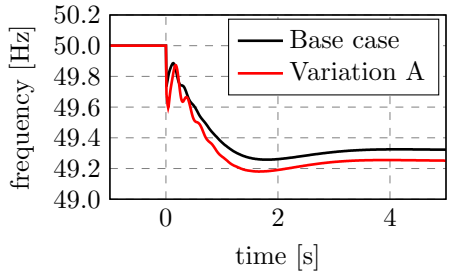}
    \caption{Electrical frequency measured at the main busbar}
    \label{fig:frequency}
\end{figure}

\begin{figure}[h!]
    \centering
    \input{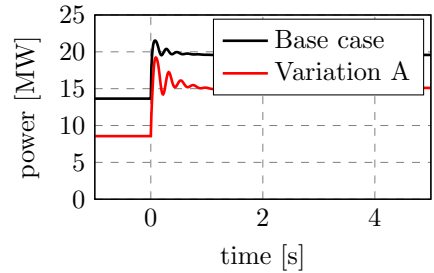}
    \caption{Active power output of the remaining gas turbine}
    \label{fig:Power}
\end{figure}

The frequency course during the simulated events is displayed in \cref{fig:frequency}, and the active power output of the remaining gas turbine (one of the two remaining gas turbines in the base case) is displayed in \cref{fig:Power}.

In the zero-wind-power simulation, around \SI{13.5}{MW} of generation is lost, and this power shortage can be handled by two remaining gas turbines that share the burden (black curves). 
In the full-wind-power simulation, only one remaining gas turbine has to handle the situation alone (red curves).
However, only around \SI{8.3}{MW} of generation is lost, because more than half of the power comes from the wind turbines, with the gas turbines operating at low power set-points.
The remaining gas turbine (also at a low power set-point) has larger headroom for handling the incident.
Both events lead to a significant frequency disturbance, which is accompanied by some decaying electro-mechanical oscillations.
The simulations show the electro-mechanical transients in case of a major power imbalance, as an example of what such an electro-mechanical RMS-domain model can simulate.

It should be noted that all of the simulations were performed with a \emph{standard} wind turbine controller.
This means operation at maximum power without upward headroom, and no response to the frequency disturbance with supportive control actions like fast frequency support or virtual inertia, which could assist the remaining gas turbine in handling the incident.
Load shedding, which also takes action to handle frequency disturbances, is not yet implemented in the simulation model.

\section{Conclusion}

This article has presented the LEOGO platform as an open reference case that can be easily shared and where no data are confidential. Its main purpose is to make it easy to investigate and compare the operation of low-emission oil\&gas platform energy systems, or more general off-grid energy systems, to test modelling concepts, operational planning strategies, control implementations etc. in a transparent way. It facilitates comparisons of results obtained using different approaches and makes it easier for modellers to benefit from each other than is the case for studies where the underlying assumptions and data are confidential or poorly documented.

The specification has a strong emphasis on the energy system, and more comprehensive descriptions of the fluid transport and processing systems  are left for potential future work. Also, a description of the reservoir would be a natural extension.
Ongoing research is investigating processing system constraints and the links between flow rates and energy demand, with the aim to develop linearised models for investigating the potential for utilising energy system flexibility in combination with variable wind. 
Updates to the electrical model may be needed to perform more detailed electrical analyses and to investigate more advanced control strategies and to check the influence on stability. Details to be realised in future revisions include the machine-side power converters and the machines of the VSDs and a power management system that coordinates the operation of the gas turbines.

As a final remark, we expect that the specification and the models presented will be revised and updated with new version tags when relevant. Contributions from others are welcome.

\section*{Acknowledgements}

The work reported here has received financial support by the Research Council of Norway through PETROSENTER LowEmission (project code 296207).
We would like thank our industry partners in the LowEmission centre for valuable feedback and input to the LEOGO specifications.

\bibliography{bibliography}

\end{document}